
\hsize29pc
\vsize42pc
\magnification=1200
\bigskip
\centerline{\bf Hydrodynamics and Nonlocal Conductivities in Vortex States}
\bigskip
\centerline{\rm Ryusuke~Ikeda**}
\smallskip
\centerline{\rm Department of Physics, Indiana University, Bloomington
       IN 47405}
\bigskip
\bigskip
\bigskip
\bigskip
\bigskip

  A hydrodynamical description for vortex states in type II superconductors
with no pinning is presented based on the time-dependent Gintzburg Landau
equation (TDGL). In contrast to the familiar extension of a single vortex
dynamics based on the force balance, our description is consistent with the
known
rotating superfluid hydrodynamics and, at long distance limit, with that
following from the Landau level approach for high field case, and enables one
to study
nonlocal conductivities perpendicular to the field in terms of Kubo
formula. Typically, the nonlocal conductivities deviate from the usual vortex
flow expressions as the nonlocality parallel to the field becomes weaker
than
that
perpendicular to the field measuring a degree of positional correlations,
and,  for instance,
the dc Hall conductivity nonlocal only in directions perpendicular to the
field becomes  zero in the vortex lattice with infinite shear viscosity.
Various
situations are discussed based on the resulting expressions.
\vfill\eject

\leftline{\bf 1. Introduction}

At present, understanding theoretically the nonlocality~[1,2] of linear
resistivity
in a clean system (with no pinning) seems to be one of central problems on the
vortex states
of a type II superconductor. By analogy to the usual viscous fluid,
Marchetti and Nelson~[1] have previously proposed within a linear
hydrodynamics that, deep in the liquid regime, a shear-viscous force
proportional to
$-\partial^2_{\perp} \partial_t{\bf s}$ will take the place of the shear term
in the elastic force ${\bf f}_{\rm el}$ in the following force-balance
 equation for the vortex lattice
 $$\partial_t(-\Gamma_1~{\bf s}+\Gamma_2~
(\hat z\times {\bf s} ))+ {\bf J}\times {\bf B} = {\bf f}_{\rm el},
\eqno (1)$$ and that consequently, a positive nonlocal term proportional to
$q^2_{\perp}$
should exist in the conductivity perpendicular to ${\bf B}$. Here
${\bf s}$ denotes the displacement field of vortex
positions, ${\bf J}$ the external current, ${\bf B}$ the uniform flux density
parallel to
 $z$ axis, $q_{\perp}$ the wavevector perpendicular to ${\bf B}$, and
 $\Gamma_1$ and $\Gamma_2$ will be given later. This equation is essentially
 an extension [3] of the single vortex dynamics to interacting vortex states.
 Subsequently, the presence of other nonlocal terms
was proposed~[2] even in the case with no sample disorder on the basis of
experiments in heavily twinned samples [4], and, deep in the liquid regime,
the presence of a large conductivity nonlocal in the field direction was
argued. It is important to {\it theoretically} clarify to what degree these
proposals can be
justified beyond the phenomenology~[1,2,5] and within a basic dynamical
model for type II superconductors such as the time dependent Gintzburg-Landau
equation(TDGL) which reasonably describes the {\it uniform} linear dissipation
in terms of Kubo formula~[6].  Actually, the eq.(1) is not correct
in the following senses. Firstly, it  cannot be used in calculations
of conductivity based on the use of Kubo formula~[5,6]. Secondly,
it is not clear from (1), in which ${\bf f}_{\rm el}$ is independent of
${\bf J}$ and of
the time derivative
$\partial_t$, how the possible viscous (nonlocal) terms in the dc conductivity
are changed as the vortex lattice is formed.
Further,  the dispersion $\sim q^4_{\perp}$ of the shear mode found as a phase
fluctuation formally and generally~[7,8] cannot be seen in eq.(1).

In the present paper, a
correct hydrodynamical approach, or equivalently, a dynamical harmonic
analysis on the transport in a vortex state, particularly the vortex lattice,
is
 presented on the basis of TDGL. Our approach has no difficulties mentioned
 above and
makes it possible to
study nonlocal conductivities above and below the melting transition with no
pinning, and  several consequences on the resulting conductivities are
discussed.  As has been understood through studies [9] on nonlocal linear
and static
responses, the use of the harmonic analysis for 3D and layered systems often
lead to misunderstandings on physics in the liquid regime, and hence we will
not
consider situations with negligibly small positional correlations of
vortices.

\smallskip
\smallskip
\leftline{\bf 2. TDGL hydrodynamics}

Our starting point is TDGL describing dynamics of the order parameter $\psi$
$$a(\gamma+{\rm i}\gamma')\partial_t\psi+[b(|\psi|^2-\rho_0)+a\xi_0^2(-{\rm i}
\partial-{2\pi\over \phi_0} A)^2]\psi=0,\eqno(2)$$ where ${\rm curl}A
=B\hat z$, $\gamma$, $a$, and $b$ are positive, $\phi_0$ is the flux quantum,
$\xi_0$ the
coherence length, and $\rho_0$ is the spatial average of the mean squared
order parameter. Since the linearized version of (2) is considered throughout
the paper, the noise term introducing fluctuations {\it in equilibrium} was
neglected, and growths of time scales will be phenomenologically included
later  as the only fluctuation effects. Within the harmonic
analysis given below,
spatial variations of the gauge field  have to be neglected in a consistent
 sense with the corresponding derivation of the infinite diamagnetic
 susceptibility in the Meissner state.

Particularly upon calculations of linear responses, it is convenient to work
rather within the {\it harmonic} approximation of the corresponding `quantum'
action with the gauge field
{\it disturbance} ${\delta}A$ (and with $\hbar=1$ )
$$S_{\rm har}=\int_r[\,\beta\sum_{\Omega}\,a(\,\gamma|\Omega|+\,{\rm i}
\gamma'
\Omega)|\psi_\Omega|^2
+\int_\tau[\,a\xi_0^2|(-{\rm i}\partial-{2\pi\over \phi_0}(A+
 {\delta}A(\tau)))\psi(\tau)|^2$$ $$+b(2|\psi_0|^2|\psi(\tau)|^2+{1\over 2}
 (\psi_0^{*2}\psi^2(\tau)+\,{\rm c.c.}))]].\eqno(3)$$
 Since
we are interested only in frequencies and wavevectors accompanying
${\delta}A$
and summations with respect them are not performed,
examinations based on (3) are
equivalent to the linearized analysis of (1) and formally independent of
temperature. Here $\tau$ is the imaginary time, ${\psi}_{\Omega}$ is the
temporal Fourier transform of ${\psi}(\tau)$,  $\Omega$ is Matsubara
frequency, $\beta$ the
inverse temperature, and $\psi_0$ the mean field solution of $\psi$.  The
superconducting (and real) part $\sigma_{ij}$ of dc linear conductivity
tensor
should be
always calculated in terms of Kubo formula
$$\sigma_{ij}({\bf q})={{\partial{\rho_s({\bf q},\,{\rm i}\Omega_0)}}\over
{\partial{\Omega_0}}}\bigg|_{{\Omega_0}{\to} +0}
={{\partial}\over {\partial{\Omega_0}}}\,{{{\delta^2}(
-\beta^{-1}{\rm ln\,Tr}_
{\psi}\,\exp(-S_{\rm har}))}\over {{\delta}A_i({\bf q},\Omega_0)\,{\delta}A_j
(-{\bf q},-\Omega_0)}}\bigg|_{{\Omega_0}{\to}+0},\eqno(4)$$
where $\Omega_0$ is the external (Matsubara) frequency, $i,\, j=x$ or $y$,
and
the absence of dc uniform (${\bf q}\!=\!0$) superfluid density was used(see
below). Since the ac and uniform case of (4) always becomes the usual vortex
flow expression in the
present formulation, the dc conductivity will be only considered hereafter.
Since
the harmonic
modes
appearing in (4) are spatial and temporal variations around the mean field
solution linearly excited by the gauge field disturbance, the linear response
quantities in vortex states resulting from such a harmonic analysis [10] are
not accompanied by the temperature $\beta^{-1}$, implying that this approach
is not applicable to the linear dissipation parallel to $B$, which is a
consequence of critical fluctuations [6]. Therefore, considerations are
largely focused on the vortex flow configuration.

Before presenting general results, it is useful to first see results on the
linear
conductivity perpendicular to ${\bf B}$ in the Landau level (LL) approach
within the lowest and next-lowest LLs (see sec.4 in Ref.7). In such a high
field approximation the
mean field solution and the shear elastic mode are found within the lowest
LL, and the
next-lowest LL,
which does not participate in constructing the mean field
solution [7, 9],  provides the uniform displacement of vortices. Since, in
constructing basis functions of eigenmodes, Matsubara frequencies in the
`quantum' action play similar roles to the wavevector parallel to ${\bf B}$
, including dynamical terms is easily performed following Ref.7. For
later convenience an amplitude-dominated mode in the lowest LL, denoted by
$\delta\rho$, will also be included. Then, to the lowest order in $q_\perp$,
the
action (3) becomes $S_{\rm LL}=S_0+S_1$, where
$$S_0{\simeq}{\beta}a{\sum_{\Omega}}\int_r
(\,\gamma\rho_0|\Omega||\chi_{\Omega}|^2
+\gamma'\Omega\,\delta\rho_{\Omega}^*\,\chi_{\Omega}
+{b\over {2a}}|\delta\rho_{\Omega}|^2$$ $$
+\rho_0{\xi_0^2}|{\partial_z}\chi_{\Omega}
-{{2\pi}\over {\phi_0}}\delta{A_{z\Omega}}|^2
+{{C_{66}}\over {2a}}\,r_B^4|{\partial_{\perp}^2}\chi_\Omega|^2),
\eqno(5)$$
$$S_1\simeq{\beta{\rho_0}a\over {2r_B^2}}\sum_{\Omega}{\int_r}
(\,\gamma|\Omega|
|{\bf s}_
{0\Omega}|^2+\gamma'\Omega({\bf s}^*_{0\Omega}\times{\bf s}_{0\Omega})_z
+{\xi_0^2}|\partial_z{\bf s}_{0\Omega}|^2$$ $$+{{2
\xi_0^2}\over {r_B^2}}|{\bf s}_{0\Omega}+B^{-1}({\delta}{\bf A}_
{\perp0\Omega}
\times
\hat z)|^2),\eqno(5)'$$
where ${\bf s}^T$(the transverse component of ${\bf s}$) is given
by $r_B^2(\partial\chi\times
\hat z)$
with vortex spacing $r_B=\sqrt{\phi_0/{2\pi B}}$, ${\bf s}_0$ the uniform
displacement with vanishing $q_{\perp}$, $\chi$ a longitudinal phase
variable,
and $C_{66}$ the
 resulting
shear modulus. Variational equations with replacement ${\rm i}\Omega\to
\omega$ give eigenmodes of the vortex lattice. Note that the resulting
dispersion $\omega$(real frequency)$=-{\rm i}({a\gamma\,\rho_0})^{-1}C_{66}\,
({q_{\perp}\,r_B})^4$ for the
 shear mode[7,8] is
different from that following from (1), and hence that the extrapolation
[1,3]
of the uniform (or equivalently, the single-) vortex dynamics to the shear
mode  is not justified within TDGL. Through the London limit (7) given below,
the
action (5) with the constraint ${\bf s}^T=r_B^2(\partial\chi\times\hat z)$
is
found to correctly give hydrodynamical results at longer distances than
$$l=r_B\sqrt{{C_{66}{r_B^2}}/{2a{\rho_0}{\xi_0^2}}},$$  which is
typically of
the order $r_B$.

 The absence of the dc uniform superfluid density (helicity modulus) is
 obvious from
(4) and $(5)'$, implying that a constant twist pitch ${\delta}A_0$ is
transmuted into an uniform displacement of vortices keeping the free energy
invariant. That is, situation is different from an elastic matter with free
energy invariant with respect to uniform displacements.
 From (4) and $(5)'$,  we easily obtain the mean field expressions on diagonal
($i=j$) and Hall ($i{\neq}j$) vortex flow conductivities
$$\sigma_{ij}(q_{\perp}=0,\,q_{\parallel})
=\rho_{s0}{{r_B^2}\over {2{\xi_0^2}}}{(\gamma\,\delta_{ij}
+\gamma'\,\varepsilon_{ij})\over {(1+{r_B^2}{q_{\parallel}^2}/2)^2}},
\eqno(6)$$
where $\rho_{s0}=2({2\pi{\xi_0}}/{\phi_0})^2{\rho_0}a$ is the mean field
superfluid density in zero field.
Due to the constraint mentioned above, (5) does not give any nonlocal (and
longitudinal) corrections to (6) associated with the freezing to the vortex
lattice. Consistently with (6),  the nonlocal superfluid density, namely
the real part of the linear response function ${\rho_s}
({\bf q}, \omega+{\rm i}0^+)$, becomes
$\rho_s(q_{\perp}\!
=\!0, q_{\parallel})\simeq\rho_{s0}{r_B^2}{q_{\parallel}^2}/2$ in dc limit.
As pointed out in Ref.9, this $q_{\parallel}^2$ behavior does not change
through a 3D melting transition [11].

 Next let us here comment on eigenmodes of (3) in 2D and nondissipative
 ($\gamma=0$) case,
corresponding to the rotating superfluid $\,^4{\rm He}$ at zero temperature
described according to the Gross-Pitaevskii equation.
Consistently with Ref.12
and 13, the order parameter in (3) will be divided into the amplitude $\rho$
and phase $\varphi$; $\psi=\sqrt\rho\,\exp({\rm i}\varphi)$ and, just as in
the
usual London approximation, the presence of the field-induced vortices will
be taken into account through the topological condition[14]
on $\partial_\mu\,\varphi=(\partial\varphi, \partial_\tau\,\varphi)$ by
neglecting any
fluctuation-induced vortices. As a result, we obtain the following harmonic
action  $$S_{\rm ph}=a\int_{r,\,\tau}\!(\,{\rm i}\gamma'\delta\rho\,\partial_
\tau\chi\,+\,\rho_0\xi_0^2(\partial\chi\,
-\,{r_B^{-2}}(\hat z\times{\bf s}))^2\,+{b\over {2a}}{\delta\rho}^2\,
+\,{C_{66}\over {2a}}(\partial_{i}s^T_j)^2).\eqno
(7)$$  The only difference of this action from (5) is that the {\it
longitudinal} current
$${\bf v}_s\,=\,2a\xi_0^2(\partial\chi\,
-\,{r_B^{-2}}({\hat z}\times\,{\bf
s}^T))\equiv\,2a\xi_0^2\,{\tilde {\bf v}}\eqno(8)$$
is nonzero in (7), which
is necessary in considering longitudinal linear responses nonlocal in
directions perpendicular to the field.  When
$q_{\perp}l>1$, the minimal coupling in (8) between $\chi$ and $s^T$ is
removed, and in principle the phase fluctuation behaves like in zero field
case at such short lengths.
Variational
equations resulting from
(7) agree with those following from an `em' analogy in Ref.12,  where an
importance of $v_s$ at nonzero frequency was stressed in a context of
superfluid $\,^4
{\rm He}$. The action (7) does not include the term
$$S_{\rm inc}
={\rm i}\int_{r,\,\tau}\!\!\gamma''{{{\rho_0}a}\over
{2\,{r_B^2}}}({\bf s}\times\partial_{\tau}{\bf s})_z,\eqno(9)$$
which is necessary in recovering the limit of incompressible fluid
(i.e., ${\delta}\rho=0$ and $v_s=0$). This addtional term with the
coefficient $\gamma''$, which should become $\gamma'$, cannot be detected
in this phase-only
analysis for (3) (Hereafter, $\gamma''$ will be assumed to be equivalent to
$\gamma'$). The eigenvalues following from $S_{\rm ph}+S_{\rm inc}$ precisely
coincide, when $\gamma'=1$, with those derived by Sonin[13] and give
dispersions $\omega_{\rm sh}^2\simeq\,
{(a\gamma')}^{-2}\,b\,C_{66}({q_\perp}\,{r_B})^4/(1
+({q_\perp}\,l_{\rm inc})^2)$  and
$\omega_{\rm c}^2\simeq\,(2{\xi_0^2}/{\gamma'{r_B^2}})^2(1
+({q_\perp}\,l_{\rm inc})^2)$ of two modes
corresponding to the shear(massless) and compression (massive) elastic modes,
respectively, where $l_{\rm inc}=\sqrt{b{\rho_0}/{2a}}\,{r_B^2}/{\xi_0}$.
The quadratic dispersion of $\omega_{\rm sh}$ at low $q_\perp$ is an origin of
the destruction [7, 8] of the long-ranged phase coherence in the vortex
lattice. The variational equation, with replacement $-{\rm i}\tau\to{t}$(real
time), with respect to the shear displacement of $S_{\rm ph}+S_{\rm inc}$
becomes the transverse part of (1) with
no external current if $\Gamma_1=0$, $\Gamma_2
=a\rho_0\gamma''2{\pi}B/{\phi_0}$,  {\it and}
 $v_s=0$. In this case the constraint $v_s=0$ does not imply that
in $(5)'$ but rather corresponds to the limit of incompressible
fluid. Following [13], this limit is valid only at shorter lengths than
$l_{\rm inc}$, which is at most of the order of several vortex spacings in a
type II superconductor near the melting line in a moderate field.

The above agreement with the well-known rotating superfluid hydrodynamics
justifies the presence of the minimal-coupling in $v_s$ between the phase
field $\chi$ and shear displacement $s^T$. Based on this, we
assume below that the $v_s^2$ term appearing in London limit(7) will be found
at the static level
by summing up many higher LLs in GL harmonic analysis around the mean
field state. When combining this with (5) and $(5)'$, we are naturally led to
invoking the following action [15] appropriate to examining nonlocal
conductivities:
$$S_{\rm nl}=\beta{\sum_{\Omega}}[\int_q[\,{\rho_0}a|\Omega|(\,\gamma\,|\chi_
{q\Omega}|^2+{\tilde \gamma}_q{{r_B^{-2}}\over 2}|{\bf s}_{q\Omega}|^2\,)
+{{C_{66}(q,|\Omega|)}\over 2}{q_\perp^2}|{\bf s}_{q\Omega}^T|^2]$$
$$+a{\xi_0^2}{\rho_0}\int_r[\,|{{\tilde
{\bf v}}_{\Omega}}\,
-\,{2\pi\over {\phi_0}}{\delta}{{\bf A}^L_{\perp\Omega}}|^2
+{r_B^{-4}}|{\bf s}_\Omega^L
+B^{-1}(\delta{\bf A}_{\perp\Omega}^T\times{\hat z})|^2$$
$$+{{r_B^{-2}}\over 2}|\partial_z{\bf s}_\Omega|^2+|
{\partial_z\chi_\Omega}-{{2\pi}\over {\phi_0}}\delta{A_{z\Omega}}|^2]]
,\eqno(10)$$
where $\delta{{\bf A}_\perp^L}\,(\delta{{\bf A}_\perp^T})$ denotes the
longitudinal (transverse) part of the gauge disturbance defined
{\it within} $x$-$y$ {\it plane}. For later
convenience the shear modulus is assumed to have  possible frequency and
wavevector dependences, and the time scale of ${\bf s}$ was
phenomenologically changed taking
account of a possibility that it may have nonlocal corrections
${\tilde \gamma}_q-\gamma>0$, due to an origin other than the freezing to the
vortex lattice [1,2] and irrelevant to statics of vortex states at long
distances.
The amplitude mode leading to a ${\Omega}^2{|\chi|^2}$ term was neglected.
The
term (9) has to be included when deriving a nonlocality of Hall conductivity.
Again, the variational equation with respect to $s^T$ of
(10) is
different from (1) with $\Gamma_1=a{\rho_0}\gamma{2}{\pi}B/{\phi_0}$
due to the
presence of nonzero $v_s$. Further, by neglecting the gauge disturbance
${\delta}A$ and examining the dispersion of the shear mode in the vortex
lattice, it is found that, due to the presence of the first term in (10),
setting $v_s=0$ from the outset, as in (1), always leads to an erroneous
result on the dispersion. Rather, at low enough $q$, the constraint
$v_s\simeq0$ is established,
and consistently, the second term in (10) becomes
unimportant compared to the first term. The first term in (10) having the
same form as in the lowest LL case (5) is required within TDGL formalism
and actually is justified because the
$\chi$ variable to be related to the shear displacement at small
but nonzero $q_{\perp}$ was shown[7] to, up to the lowest order in
$q_{\perp}$, become a phase change around $\psi_0$ even if higher LLs
are {\it fully} included.

\smallskip
\smallskip
\leftline{\bf 3. Nonlocal Conductivities}

It is straightforward to, using (4) and (10), find the nonlocal
 conductivities. First, let us discuss the vortex lattice
 where $C_{66}(q=0, \Omega=0)\ne0$. In this case the longitudinal
 ($\parallel {\bf q}_\perp$) part of the diagonal
 conductivity and the Hall conductivity are given by  $$\sigma_{xx}^{(s)L}(q)
 =\rho_{s0}{{r_B^2}\over {2{\xi_0^2}}}\,{{{{\tilde \gamma}_q}{q_\parallel^4}
 +\gamma{q_\perp^4}l^2(q_\parallel^2+2(l{q_\perp}/r_B)^2)}\over
 {(q_\parallel^2(1+{q^2}{r_B^2}/2)+l^2q_\perp^4)^2}},$$  $$\sigma_{xy}^{(s)}(q)
 =\rho_{s0}{{r_B^2}\over {2{\xi_0^2}}}\,{{\gamma'{q_\parallel^2}}\over
 {(1+{q_\parallel^2}{r_B^2}/2)({q_\parallel^2}(1+{q^2}{r_B^2}/2)
 +{q_\perp^4}{l^2})}}.\eqno(11)$$
 using the length $l\propto{\sqrt{C_{66}}}$ defined previously. The
 transverse part $\sigma_{xx}^{(s)T}$ of the diagonal conductivity merely
 becomes, even in the liquid regime, ${{\tilde \gamma}_q}\sigma_{xx}({q_\perp}
 =0, q_\parallel)/{\gamma}$ (see (6)), and hence, in contrast to (11), is not
 affected by the positional correlation [16]. We note that the term
 $\sim
 {q_\parallel^2}+{{q_\perp^4}l^2}$ in denominators is the dispersion of the
 3D shear
 (Goldstone) mode destroying the true off-diagonal long ranged order [7,8].
 In general, the magnitudes of these conductivities significantly
 depend much on the {\it relative} size of $|q_\parallel|$ and $|q_\perp|$.
 For  instance,
when $|q_\parallel|>l|q_\perp|/{r_B}\sim|q_\perp|$, both $\sigma_{xx}^{(s)L}$
and $\sigma_{xy}^{(s)}$ are well approximated by $\sigma_{ij}$ given in (6)
with replacement $\gamma\to{\tilde \gamma}_q$. On the contrary, when
$|q_\parallel|<l{q_\perp^2}$, $\sigma_{xx}^{(s)L}$ typically becomes the value
 $\gamma{\rho_{s0}}/{{q_\perp^2}{\xi_0^2}}$ at zero $q_\parallel$
 (or in 2D case).
  This expression is the same as the 2D result in zero field [17](see also
  [5]), reflecting the fact that it arises entirely from the phase variation
  $\chi$.
  Correspondingly,
 $\sigma_
{xy}^{(s)}$ decreases like $\sim({q_\parallel}/{q_\perp^2})^2$ and
{\it vanishes}
at zero $q_\parallel$ or in 2D case. In
 other words,
  as the nonlocality parallel to $B$ becomes negligible compared to that
  perpendicular to $B$, contributions of a finite ${\tilde \gamma}_q$ are
  covered and overcome by
  the perfect positional correlation, i.e., the presence of the Goldstone
  mode. This feature is lost, at fixed $q_\perp$ and $q_\parallel$,
  as $l$ becomes shorter, i.e., with increasing
  field, which is consistent with the absence of the nonlocal conductivity
  induced by the positional correlation within the lowest LL (see a sentence
  following (6)). In addition, note that, as seen in ${q_\parallel}\to0$
  limit, the infinite shear viscosity of the vortex lattice does not mean a
  zero
  value
  of the nonlocal resistivity. We emphasize that the remarkable difference
  between the uniform (${q_\perp}=0$) result (6) and this large conductivity
  in the case with zero $q_\parallel$ and
  small $q_\perp$ is a consequence of the fact that, at zero $q_\perp$, the
  shear displacement $s^T$ decouples with the zero mode
  $\chi$ and changes into the uniform displacement, and hence that it is
  theoretically correct.

  The above result that
  $\sigma_{xx}^{(s)L}\sim
  {q_\perp^{-2}}$ and $\sigma_{xy}^{(s)}=0$ when $q_\parallel=0$ and
  ${q_\perp}\ne0$ is  also applicable to a state with nonzero
  $C_{66}(q\ne0, |\Omega|=0)$ such as the 2D hexatic liquid state which may
  be possible in low enough fields (The 3D hexatic phase [1, 10] is
  inconsistent with the first order freezing to the vortex lattice and not
  possible in real superconductors, and hence need not be considered). It is
  valuable to examine these dc linear responses in real experiments in
  relation to determinations of the position of the freezing to the vortex
  lattice and of the existence or absence of the hexatic phase in 2D systems.

  When trying to understand the liquid regime (disordered state) in the
  present approach, some comments are necessary. As pointed out in Ref.9, the
  harmonic analysis cannot be used in the situation where the contributions
  with nonzero reciprocal lattice vectors are negligible even if the amplitude
  fluctuation of the order parameter is negligible. In such situations
  the nonvanishing transverse diamagnetic susceptibility results from the
  static
  superconducting fluctuation and, in layered systems, shows a dimensional
  crossover due to a competition between the layer spacing and a finite phase
  coherence length which cannot be found in the harmonic analysis.
  Consistently, this dimensional crossover will be seen
  also in
  the nonlocal conductivity.  On the other hand, it is practically difficult
  at present to
  understand possible nonlocal corrections,
  just above the freezing point to the vortex lattice, to the vortex flow
  expression (6) according to the fluctuation theory [6]. For these reasons,
  we will only consider the case just above the freezing point to the vortex
  lattice and with
  vanishing $q_{\parallel}$ (see, however,
  the last section).

  In this case,  the corresponding results to (11) are given by
  $$\sigma_{xx}^{(l)L}({q_\perp}, {q_\parallel}=0)=\rho_{s0}{{r_B^2}\over
  {2{\xi_0^2}}}\,{{{{\tilde \gamma}_q}
  +\eta{q_\perp^2}{r_B^2}/{{\rho_0}a}}\over
  {1+{q_\perp^2}{r_B^2}({{\tilde \gamma}_q}
  +\eta{q_\perp^2}{r_B^2}/{{\rho_0}a})/{2\gamma}}},$$
  $$\sigma_{xy}^{(l)}({q_\perp}, {q_\parallel}=0)
  =\rho_{s0}{{r_B^2}\over {2{\xi_0^2}}}\,{{\gamma'}\over
  {1+{q_\perp^2}{r_B^2}({{\tilde \gamma}_q}
  +\eta{q_\perp^2}{r_B^2}/{{\rho_0}a})/{2\gamma}}},\eqno(12)$$
  where the Maxwell form [18] $C_{66}(q,|\Omega|)\simeq\eta|\Omega|$
  with shear
  viscosity $\eta$ for the dynamical shear modulus was assumed.
  Interestingly, the expressions (12) {\it smoothly} lead to (11) in zero
  $q_\parallel$ for the vortex lattice
 by taking $\eta$ to be infinity. The terms $1
 +\eta({q_\perp}{r_B})^4/{2\gamma{\rho_0}a}$
 in the denominators of (12) again originates from the spectrum
 $\omega\sim\,-{\rm
 i}C_{66}({q_\perp}{r_B})^4$ of the shear mode, which cannot be found
 from (1). Consequently,
 at high $q_\perp$ of the order ${r_B}^{-1}$,
 both
 (11) and (12) are well approximated by (6), suggesting that, for such a
 rapid
 variation of the external current, not only collective effects but also the
 sample disorder is irrelevant because
 the time scale to be affected by the sample disorder (pinning) will be
 ${\tilde \gamma}_q$ and not be $\gamma$. In general, in the case with a
 pinning effect where ${\tilde \gamma}_q$ will grow, the viscous effect
 accompanying the freezing transition becomes relatively negligible. Further,
 the expressions (12) suggest that the low temperature limits of the nonlocal
 conductivities for small $q_\perp$ in 2D systems with pinnings, where we have
 no
 transition at nonzero temperatures, again become the corresponding ( and
 above-mentioned ) results in the vortex lattice independent of
 ${\tilde \gamma}_q$.

\smallskip
\smallskip
\leftline{\bf 4. Comments and Conclusion}

Finally, we will discuss consequences of possible viscous effects [1, 2] in
pinning-free systems, {\it other than} $\eta$, which should appear in
nonlocalities of ${\tilde \gamma}_q$. According to (11) and (12), when
$|q_\parallel|\ll|q_\perp|$, it is difficult to
practically divide contributions of ${\tilde \gamma}_q$ from the positional
correlation, and hence, we will only consider the case with vanishing
$q_\perp$ but nonzero $q_\parallel$, where the diagonal conductivities can be
always expressed by (6) with replacement $\gamma\to{\tilde \gamma}_q$.
Further we will focus on the 3D region below a dimensional crossover suggested
in sec.3 (We note  that this dimensional crossover is quite different
from that argued through experiments in Ref.4). For
this case, it is often argued [19]
that
thermally activated cutting (and reconnection) processes among vortices
induced by a nonuniform (${q_\parallel}\ne0$) current will lead to a very
long length scale and significantly increase ${\tilde \gamma}_q$ even if the
(thermally-induced) entanglement is absent [2]. This picture suggests that
the time
scale grows unlimitedly even in the vortex lattice under cooling, and
hence that the bent vortices cannot move at low enough temperatures. It should
be noted that
 such an argument is also applicable to a nondissipative case
 (with zero $\gamma$ but nonzero $\gamma'$) by imagining a nonlocal
growth of $\gamma'$. However, it is even unclear to us if this is a correct
argument in the context of the {\it linear} hydrodynamics. In superfluid
$^4{\rm He}$[20]
the reconnection process does not seem to need a remarkable slow dynamics.
In
addition, the system-size dependence of a temperature [4, 21] characteristic
of apparently nonlocal vortex motions is inconsistent with the argument
based on
the thermally activated vortex cutting processes, because the measured size
dependence inevitably means an algebraic decrease of the cutting barrier with
increasing the system-size, although intuitively such a decrease of the
barrier should not be expected.
In relation to this issue, we emphasize that viscous terms appearing in
${\tilde \gamma}_q$ cannot reduce to elastic (and static) terms in the vortex
lattice and that the (if any) entanglement [1] making a growth of
${\tilde \gamma}_q$ possible must disappear deep in the vortex lattice
[22]. In Ref.9, the presence of a large and positive nonlocal ($\sim{q^2_z}$)
contribution
in ${\tilde \gamma}_q$ was questioned on the basis of the observation that,
in contrast to the pinning-induced activation form of the time scale,
 any nonlocal (i.e., wavevector-dependent) growth of the time scale
may be incompatible with the uniform vortex flow due to the nonlinearity in
Kubo formula in the nonGaussian fluctuation theory [6]. Nevertheless,
a
possibility of a large ${\tilde \gamma}_q$ for nonzero $q$ should be
  further searched in an extention of the fluctuation theory [6]
  which is not available at present. Experimentally, measurements in twin-free
  samples
  corresponding to those
in [4, 21] should be performed.  Actually, the data on the resistivity
parallel to $B$ in [21] seem to intuitively contradict  the data in a more
3D-like situation [23].

In conclusion, a hydrodynamics for vortex states in type II
superconductors consistent with the rotating superfluid hydrodynamics has been
presented in order to study
the nonlocal conductivities perpendicular to the applied field. In
particular,
the
vanishing Hall conductivity and divergent diagonal conductivity in the case
with nonlocality in perpendicular directions to the field
 will be applicable to (if any) 2D hexatic state and
useful in experimentally judging its existence or absence in type II
superconductors.

I acknowledge related discussions with Alan Dorsey, Wai Kwok, Chung-Yu Mou,
and
Ulrich Welp. This research was finantially supported by `Kyoto University
Foundation' and by DOE under Grant No. DE-FG02-90ER45427 in Indiana
University.
\bigskip
\bigskip
\frenchspacing
** On leave from  Department of Physics, Kyoto University, Kyoto 606, Japan
\item{[1]}  M.C.Marchetti and D.R.Nelson, Phys.Rev.B {\bf 42}, 9938(1990).
\item{[2]}  D.A.Huse and S.N.Majumdar, Phys.Rev.Lett. {\bf 71}, 2473(1993).
\item{[3]}  For instance, L.P.Gor'kov and N.B.Kopnin, JETP{\bf 33}, 1251(1971).
\item{[4]}  H.Safar et al., Phys.Rev.Lett. {\bf 72}, 1272(1994).
\item{[5]}  C.Y.Mou, R.Wortis, A.T.Dorsey, and D.A.Huse, preprint.
\item{[6]}  R.Ikeda, T.Ohmi, and T.Tsuneto, J.Phys.Soc.Jpn. {\bf 60}, 1051
(1991); S.Ullah and A.T. Dorsey, Phys.Rev.B {\bf 44}, 262(1991).
\item{[7]}  R.Ikeda, T.Ohmi, and T.Tsuneto, J.Phys.Soc.Jpn. {\bf 61}, 254
(1992); ibid {\bf 59}, 1740(1990).
\item{[8]}  M.A.Moore, Phys.Rev.B {\bf 45}, 7336(1992); G.Baym, preprint.
\item{[9]}  R.Ikeda, J.Phys.Soc.Jpn.(to be published).
\item{[10]}  T.Chen and S.Teitel, Phys.Rev.Lett. {\bf 72}, 2085(1994). As done
by these authors, imposing the London gauge for the gauge {\it disturbance}
in calculations of linear responses is misleading and has no theoretical
foundation. The absence of the longitudinal component in the superfluid
density at {\it zero frequency} in the Meissner state has nothing to do
with wavevector dependences of the gauge disturbance.
\item{[11]} This means that the elastic terms associated with the compression
mode are not affected by the 3D melting transition (see [9]). It also
invalidates a corresponding result $\sim|q_\parallel|$ in a putative liquid
phase [Feigel'man and Ioffe,  preprint] even if the fluctuating magnetic
field is neglected, because it would mean that such an intermediate phase has
a stronger ordering than in the vortex lattice.
\item{[12]} S.C.Zhang, Phys.Rev.Lett.{\bf 71}, 2142(1993).
\item{[13]} See (4.71) to (4.73) in E.B.Sonin, Rev.Mod.Phys.{\bf 59},
87(1987). Parameters $c_T$, $c$,
and $\Omega$ appearing in this reference are given using our notation by
$c_T^2\equiv{2{\xi_0^2}C_{66}}/{a{\rho_0}{\gamma'^2}}$, $\Omega\equiv{\xi_0^2}
/{{r_B^2}\gamma'}$, and $c^2\equiv{2b{\rho_0}{\xi_0^2}}/{a{\gamma'}^2}$.
\item{[14]} See Appendix D in [7], and  V.N.Popov, {\sl Functional integrals
in Quantum Field Theory and Statistical Physics}(Reidel, Dordrecht, 1983).
\item{[15]} See also Appendix D in [7], and A.I.Larkin and Y.N.Ovchinikov,
J.Low Temp.Phys. {\bf 34}, 409(1979).
\item{[16]} Correspondingly, we have two kinds of the transverse parts of
diamagnetic susceptibility due to the presence of two (elastic) modes, and
they are given by $\chi_\perp^{(s)}\!={r_B^2}{\rho_{s0}}[{q_\parallel^2}
+2(l{q_\perp}/{r_B})^2]/[{q_\parallel^2}(2+{(q{r_B})^2})
+2{l^2}{q_\perp^4}]$ and $\chi_\perp^{(c)}\simeq{r_B^2}{\rho_{s0}}/2$ for
the shear and compression modes, respectively. Note that the divergent
behavior of ${\chi_\perp^{(s)}}({q_\parallel}=0, {q_\perp}\to0)$ does not
contradict the vanishing superfluid density perpendicular to $B$. Since we
have neglected in $(5)'$ and (10) an O$({q_\perp^2}{s^{L2}})$ term playing
no roles in nonlocal linear dissipations, the longitudinal susceptibility is
zero in the present approach.
\item{[17]} H.J.Mikeska and H.Schmidt, J. Low Temp. Phys. {\bf 2}, 371(1970).
\item{[18]} For instance, V.M.Vinokur et al., JETP {\bf 73}, 610(1991).
\item{[19]} For instance, C.Carrano and D.S.Fisher, Phys.Rev.B {\bf 51}, 534
(1995).
\item{[20]} K.W. Schwartz, Phys.Rev.B {\bf 38}, 2398(1988).
\item{[21]} D.Lopez et al., Phys.Rev.B {\bf 50}, 7219(1994).
\item{[22]} In addition, the picture in [1] that the thermally-induced
entanglement will become more remarkable with increasing field is
questionable, because
 a situation dominated by the lowest LL mode, in which the entanglement is
 absent (see [9] and M.J.W. Dodgson and M.A.Moore, preprint), must become
 valid with
increasing field.
\item{[23]} W.K.Kwok et al., Phys.Rev.Lett. {\bf 64}, 966(1990).

\bye